# The Medieval Inquisition: Scale-free Networks and the Suppression of Heresy

**Paul Ormerod** (pormerod@volterra.co.uk)

Volterra Consulting

**and**

**Andrew P. Roach (**A.Roach@history.arts.gla.ac.uk**)\***

Department of History, University of Glasgow

May 2003

\* Corresponding author

## *Abstract*

*Qualitative evidence suggests that heresy within the medieval Catholic Church had many of the characteristics of a scale-free network. From the perspective of the Church, heresy can be seen as a virus. The virus persisted for long periods of time, breaking out again even when the Church believed it to have been eradicated. A principal mechanism of heresy was through a small number of individuals with very large numbers of social contacts.*

*Initial attempts by the Inquisition to suppress the virus by general persecution, or even mass slaughtering, of populations thought to harbour the 'disease' failed. Gradually, however, the Inquisition learned about the nature of the social networks by which heresy both spread and persisted. Eventually, a policy of targeting key individuals was implemented, which proved to be much more successful.*

1.   **Introduction**

There is a great deal of current interest in the dynamics of epidemics. The classical approach of the Susceptible-Infected-Recovered model of epidemiology (for example, [1]) assumes implicitly that all agents have an equal probability of meeting any other agent. It is well known that in such models there is in general a critical threshold of the proportion of agents which is infected. Below this threshold, a virus will not spread throughout the system.

Scale-free networks have recently been shown to be prone to the spreading and persistence of infection no matter how small the spreading rate [2,3]. In these networks, a small number of agents have a very large number of connections to others, and most agents have very few. An important feature of scale-free networks is that viruses both persist for much longer, and are harder to eradicate, than is expected from conventional epidemiology models [4]. Once a virus appears in the system, anti-virus strategies require targeting the highly connected few rather than the much less connected many [5].

Important social networks such as the world-wide web [for example, 6] and sexual contacts [7] have been shown to have a structure which approximates a scale-free network, though the possibility exists of exponential truncation of such power law behaviour [8].

The medieval Catholic Church in Western Europe, and particularly in France, faced strong and persistent outbreaks of heresy. The Church undoubtedly regarded heresy as a disease. This paper presents qualitative evidence that the Catholic Inquisition, a body charged with the eradication of heresy, learned to behave as if it knew that the spread of heresy in medieval society did so on a scale-free network.

## 2.     Heresy in 13th and 14th century France

The chronicler of the Albigensian Crusade of 1209, a crude attempt at the extirpation of religious dissent in the south of France by armed force, shows quite clearly that medieval observers were well aware of the parallels with disease [9]:

> 'Just as one bunch of grapes can take on a sickly colour from the
> aspect of its neighbour, and in the fields the scab of one sheep
> or the mange of one pig destroys an entire herd,' so, following
> the example of Toulouse, neighbouring towns and villages
> in which heresiarchs had put down their roots were caught up in
> the shoots put out by that city's unbelief, and became infected
> with the dreadful plague.'

The Albigensian Crusade involved general intimidation, up to the point of mass killing, of populations believed to be infected with the heresy. The policy of suppression was essentially random, with individuals and groups singled out and punished. The problem was famously recognised in the quote attributed to the papal legate and monk, Arnaud Aimeric at the storming of Béziers [10]:

> Knowing from the confessions of these Catholics that they were
> mixed up with heretics, [the crusaders] said to the abbot.
> 'What shall we do, lord? We cannot tell the good from the bad.
> The abbot, ……is said to have said: "Kill them. For God knows
> who are his." Thus innumerable persons were killed in that city.'

By the time the account was written in the early 1220s, then the futility of indiscriminate ferocity in extirpating heresy was acknowledged. The reason for its failure appears to be that the structure of the social network had many of the characteristics of a scale-free network.

For example:

- despite the fact that in the whole of Europe there were only three universities worthy of the name, ideas travelled remarkably quickly via the network of educated clerks [11].
- a small number of individuals exercised a disproportionate influence in the spread of ideas [12].
- heresy tended to linger for very long periods of time. For example, Catholic writers preparing reports for the 1274 Council of Lyon thought the threat was over [13], yet the last Cathar was only burnt in 1321 and a Cathar revival led by only ten *perfecti* around 1300 caused a major panic among Catholic churchmen. [14].

As expertise on heresy grew in the thirteenth century there would seem to be growing evidence that expert observers recognised that what they were dealing with was a scale-free network. The fight against heresy instituted the use of prison as a punishment (as opposed to a holding tank for suspects "on remand") and the 1229 Council of Toulouse stated [15]:

> Heretics…who return to Catholic unity…not…voluntarily
> are to be imprisoned by the bishop of the place….to prevent
> their having the power of corrupting others.

Medieval prison building was inevitably small scale and there was no question of using it other than for the containment of important individuals, particularly those who might be able to provide further information later.

The next question is then to what extent 'heresy hunters' targeted the highly connected individuals. In the twelfth century ,hardly at all, but after 1231 with specialist inquisitors starting to operate there was a growing body of expertise on how heresy spread and how it might be stopped. This knowledge was summarised from around 1250 in a number of handbooks for inquisitors. The most well known of these was by the Dominican friar, Bernard Gui (the name was used by Umberto Eco for the caricature of an inquisitor in *The Name of the Rose*), the *Practica Inquisitionis*, completed in 1323-24 from which most of the following is taken. Gui was interested in the connections of heretical sects; in the section dealing with Cathars he suggests suspects be asked [16]

> Whether he had any familiar association with heretics; when; how;
> And who was responsible for it.

As well as how the network was physically organised;

> Whether he received any heretical person or persons in his home;
> Who they were; who brought them there;….who visited them there
> and escorted them thence.

Although Bernard also asked what went on in houses as regards preaching or ceremonies, he was not at all interested in beliefs. Instead the guides and messengers were targeted.

Gui was also well informed about how heresy spread, in talking of the Waldensians (an evangelical movement outside the Church) he describes the activities of their spiritual elite or Perfect, having recognised their high mobility 'fleeing from city to city' he describes a visit to a local community [17]:

> When they have come to a place, word of their arrival is put
> about and many gather at their lodging place to hear and see
> them. People send whatever they have in the way of food
> and drink.

Gui was able to see that such mobility could only be achieved through considerable wealth and that the Waldensians could pass as merchants if need be [18]:

> Also, every year they hold or celebrate one or two chapters-general in some important community, as secretly as possible, gathering, as if they were merchants, in a house leased long before by one or more of their believers.

It then remains to see how this knowledge might have translated into action on the ground. The weapons the inquisitors had to deal with these highly connected individuals were the penances set to 'repentant' heretics and their supporters. A group of 224 of these have been analysed from Gourdon, north of Cahors in France in 1241-42 [19]. At first, inquisitors used pilgrimage as a punishment, related to the sentence of exile in previous centuries: 68.7% of the Gourdon penitents were packed off on pilgrimages as wide ranging as Canterbury, Rome and Constantinople as well as a variety of French shrines. But as early as 1243-44 observers noted the problems of penances involving travel 'lest through [the penitents'] perfidy the dam of faith be breached' [20], by the end of the century penitential pilgrimage was being dismissed as a temptation to misbehaviour because of the many contacts they would inevitably make across a broad geographical area [21].

There was also an attempt to mop up the wealth of well-heeled heretical supporters. Simple fiscal penalties were not regarded as sufficiently morally improving so penitents were ordered to pay for the upkeep of a number of paupers for terms ranging from one year to life, a variation made four penitents responsible for the support of priests. Some 15.7% of cases in Gourdon had this penalty imposed on them, but despite its appropriateness as a redirection of charity which had previously gone to heretics, the penance rapidly fell into disuse. It seems to have been too discreet, in that there was little element of public recognition of wrongdoing.

Attention turned instead to punishments which restricted movement or marked the penitent out, making social intercourse difficult. None of the Gourdon heretics were imprisoned, probably because the inquisitors felt too politically insecure to enforce the sentence, but in 1246, 23 from 207 suspects (11.1%) were sentenced to imprisonment in newly built jails.  By the early years of the fourteenth century in nearby Toulouse, this had risen to over 60%[22].  Imprisonment was always tilted to the punishment of the wealthy, after all prisoners had to pay for their own upkeep which could become difficult since jail was usually accompanied by the confiscation of personal wealth.

Meanwhile inquisitors turned to the problem of 'innoculating' society against the few highly connected individuals.  The penance of 'cross wearing', having two yellow crosses sown on the back and front of all visible clothing, is resonant of twentieth century totalitarian oppression, but probably started as a genuinely penitential act initiated by Saint Dominic himself at the turn of the thirteenth century [23].  In Gourdon, this was a common sentence for supporters of heretics, just over a third, (34.3%) were given it, often in conjunction with other punishments [24].  In the intensely repressive atmosphere of Languedoc in the 1240s the social implications of cross wearing quickly became disastrous.  By 1246 it became necessary to forbid people to ridicule cross wearers or refuse to do business with them. This penalty was particularly disruptive to the networks of supporters of the highly connected individual heretical preachers.  To be seen consorting with a known heretic, while wearing a cross was a sure sign that the 'penitent' was insincere and laid him or her open to more severe punishments including imprisonment and burning.  The gathering of individuals to listen to heretical preaching now became almost impossible as just to be seen with a cross wearer was to risk the accusation of heretical sympathies oneself [25, 26].

Finally, over the course of two decades or more the inquisitors developed a network of spies and agents, who often targeted key heretical figures. An example from 1293 to capture a Milanese heretic, William, on the run in the remote lands of what is now Slovenia, demonstrates how professional this operation had become.  A spy was sent by inquisitors based in Pavia into the lands east of Venice to find out where the heretic was staying.  Then a task force was assembled to apprehend William consisting of 'hunters' from the Franciscan friars.  The heretic was brought back into Italy, staying in various prisons en route, tried and finally burnt in his home town of Milan for maximum deterrent effect [27]. The cost of the operation to arrest William was 25 *libri imperiales* or £25-30 000 in modern terms [28]. But by then the inquisitors knew the value of securing highly mobile, connected individuals.

In conclusion, given the limitations of technology and the analytical tools at their disposal there is little doubt that inquisitors were aware of what we would call 'scale-free' networks and in the period 1231-1300 devised strategies to disrupt them. The legacy of these results was mixed; corruption or inquisitors' self-interest sucked innocent men and women into the machine, on the other hand there was no further outbreak of organised religious dissent in France or Italy until the Reformation.

3.     **Conclusion**

The evidence from medieval history is inevitably far more qualitative than quantitative, and is not open to the kind of rigorous analysis which can be carried out on, for example, the properties of the world wide web.

However, the evidence does suggest that medieval heresy, regarded as a disease by the Catholic Church, existed on what appears to have been an approximately scale-free network. Heresy could persist for long periods of time, and erupted again even when the Church believed it had been eradicated. Despite the popular view of medieval Europe as a stagnant society, ideas could and did travel rapidly. Further, a small number of people exercised a disproportionate influence on the spread of ideas.

Initially, the Inquisition adopted a policy of what might be thought of as random inoculation. Intimidation and even mass slaughtering of 'infected' populations were tried as policies, principally in the Albigensian Crusade in the early 13[th] century. Such policies, however, were unsuccessful.

Gradually, the Inquisition learned to adopt a much more subtle strategy, obtaining information on the nature of the network across which heresy spread, and targeting individuals with high levels of social contact. These polices proved much more effective.

**References**

[1]     J.D.Murray, *Mathematical Biology*, Springer-Verlag, Berlin, London (1990)

[2]     R.Pastor-Satorra and A.Vespignani, 'Epidemic spreading in scale-free networks', *Phys. Rev. Lett*. 86, 3200-3203 (2001).

[3]     D.S.Callaway, M.E.J.Newman, S.H.Strogatz, and D.J.Watts, 'Network robustness and fragility: Percolation on random graphs', *Phys. Rev. Lett*. 85, 5468-5471 (2000).

[4]     R.M.May and A.L.Lloyd, 'Infection dynamics on scale-free networks', *Phys. Rev. E* 64, 066112 (2001).

[5]     R.Pastor-Satorras and A.Vespignani, 'Immunization of complex networks', *Phys. Rev. E* 65, 036104 (2002).

[6]     A-L.Barabási, R.Albert, and H.Jeong, 'Scale-free characteristics of random networks: The topology of the World Wide Web', *Physica A* 281, 69-77 (2000).

[7]     F.Liljeros, C.R.Edling, L.A.N.Amaral, H.E.Stanley, and Y.Aberg, 'The web of human sexual contacts', *Nature* 411, 907-908 (2001).

[8]     S.Mossa, M.Barthelemy, H.E.Stanley, and L. A. Nunes Amaral, 'Truncation of power law behavior in "scale-free" network models due to information filtering', cond-mat/0201421 (2002)

[9] Peter of Vaux-de-Cernay, *History of the Albigensian crusade*, trans. W.A. and M.D. Sibly, Boydell, Woodbridge, 1998. The quotation is from Juvenal's *Satires*, II, 79-81.

[10] M. Barber, *The Cathars: Dualist Heretics in Languedoc in the High Middle Ages*, Pearson, Harlow, 2000, pp.211-12 n.20, quoting Caesarius of Heisterbach, *Dialogus Miraculorum*, ed. J.Strange, vol.1, Cologne, Bonn and Brussels, 1851.

[11] R.W. Southern, *Scholastic Humanism and the Unification of Europe*: vol. 1: *Foundations*, Blackwell, 1995, pp.132-45.

[12] R.I. Moore, *First European Revolution, c.970-1215*, Blackwell, Oxford, 2000.

[13] Report of Humbert of Romans to the Council of Lyon, printed as 'De corrigendis in ecclesia Latinorum' in *Sanctorum Conciliorum et decretorum Collectio Nova: Supplementum ad concilia* eds. Veneto & Labbeana, Lucca, 1749, vol. 3, cols. 1-28.

[14] Barber, *The Cathars,* pp.176-90, 199.

[15] Canon II, cited in A. P. Roach, 'Penance and the making of the inquisition in Languedoc', *Journal of Ecclesiastical History*, 52, (2001), p.427.

[16] Bernard Gui, *Practica inquisitionis heretice pravitatis*, Pt. V trans. in W. L. Wakefield & A. P.Evans, *Heresies of the High Middle Ages*, Columbia University Press, NY., 1969, p.385.

[17] ibid. p. 395.

[18] ibid. p. 393.

[19] Roach, 'Penance' *Journal of Ecclesiastical History*, 52 (2001), pp. 416-26. The cases themselves are recorded in the Fond Doat, Bibliothèque Nationale, Paris Vol.21[Bodleian Library, Oxford, MS. Film 1692], fos. 185v-213v.

[20] G. Mansi, *Nova et amplissima collectio sacrorum conciliorum*, Venice, 1759-1927, vol.23, pp.356-7.

[21] M. Mansfield, *Humiliation of sinners: public penance in thirteenth-century France,* Cornell University Press, Ithaca, NY, 1995, p.125.

[22] A. Pales-Gobilliard, 'Pénalités inquisitoriales au XIVe siècle' in *Crises et réformes dans l'église de la réforme Grégorienne à la préréforme,* Comité des travaux historiques et scientifiques, Paris, 1991, pp.143-57.

[23] Trans. in R. Brooke, *Coming of the friars*, London, 1975, p.185 and see A. P. Roach, *The Devil's World; Heresy and society, 1100-1300*, Pearson, Harlow (forthcoming).

[24] Roach, 'Penance', *Journal of Ecclesiastical History,* 52 (2001), p.422.

[25] J. B. Given, *Inquisition and Medieval Society*, Cornell University, Ithaca, 1997 p.85.

[26] J. H. Arnold, *Inquisition and Power*, University of Pennsylvania, Philadelphia, 2001, p.63.

[27] Inquisition expense accounts, Vatican 'Collectoriae' vol.133, partially published in G. Biscaro, 'Inquisitori ed eretici Lombardi, 1292-1318', *Miscellanea storia italiana*, 50 (1922), p.509.

[28] Based on 18 *libri imperiales* being a skilled labourer's annual wage in 1300, £20 000 in 2002.